\documentclass[pra,twocolumn,superscriptaddress,showpacs,floatfix,longbibliography]{revtex4-2}
\usepackage{mathrsfs,braket}
\usepackage{amssymb, amsbsy, amsmath, latexsym, dsfont, array, layout,
graphicx,mathrsfs,color,ulem,bm}
\usepackage[colorlinks=true,citecolor=blue,urlcolor=blue]{hyperref}

\begin{document}

\title{Topology and its detection in a dissipative Aharonov-Bohm chain}

\author{Haowei Li}
\affiliation{CAS Key Laboratory of Quantum Information, University of Science and Technology of China, Hefei 230026, China}
\author{Wei Yi}
\email{wyiz@ustc.edu.cn}
\affiliation{CAS Key Laboratory of Quantum Information, University of Science and Technology of China, Hefei 230026, China}
\affiliation{CAS Center For Excellence in Quantum Information and Quantum Physics, Hefei 230026, China}

\begin{abstract}
In a recent experiment, a dissipative Aharaonov-Bohm (AB) chain was implemented in the momentum space of a Bose-Einstein condensate.
Formed by a series of dissipative AB rings threaded by synthetic magnetic flux, the chain exhibits the non-Hermitian skin effect, necessitating the non-Bloch band theory to account for its topology. In this work, we systematically characterize topological features of the dissipative AB chain, particularly beyond the experimentally realized parameter regime. Further, we show that an atom-injection spectroscopy is not only capable of revealing topological edge states, as has been demonstrated in the experiment, but also the general band structure of the system. We then discuss alternative dynamic detection schemes for the topological edge states.
Given the generality of the model and the detection schemes, our work is helpful to future study of topological models with non-Hermitian skin effects across a variety of quantum simulators.
\end{abstract}

\maketitle

\section{Introduction}

The state-of-the-art quantum control in systems such as photonics~\cite{photonics1,photonics3,photonics4}, cold atoms~\cite{luole,bryce,yan,NHSOCexp}, or trapped ions~\cite{trappedion3,chenion,trappedion4} offer unprecedented access to the rich dynamics and exotic phenomena in open quantum systems that undergo particle or energy exchange with their environment.
A non-Hermitian description applies therein, for instance, by imposing post selection~\cite{Non1,Uedareview,molmer,michael,weimer}, or by mapping the density-matrix dynamics to an enlarged Hilbert space~\cite{mastereqeff1,mastereqeff2,zhushiliang,tianyu}.
The resulting non-Hermitian physics provides an unconventional perspective of open systems, and has attracted extensive interest in recent years.
Dictated by a non-Hermitian effective Hamiltonian, exotic spectral or dynamic properties, such as the parity-time symmetry~\cite{PT1,photonics2}, enhanced sensing~\cite{sensor1,sensor2,sensor3} and topological transfer~\cite{photonics4,NHSOCexp}, non-Hermitian topology~\cite{nhtopot1,nhtopot2,nhtopot3,nhtopoe1,nhtopoe15,nhtopoe16} and so on, have been systematically studied and experimentally confirmed in a wide range of physical systems.

\begin{figure}[tbp]
	\includegraphics[width=8cm]{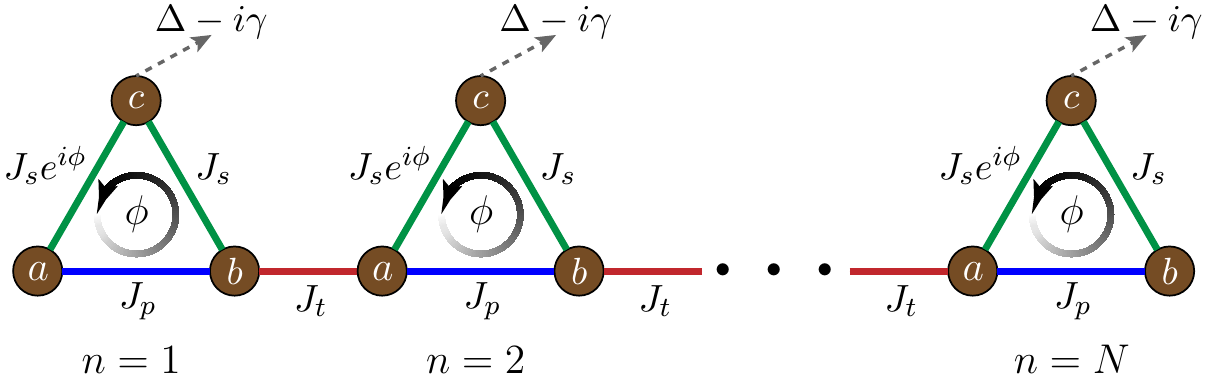}
	\caption{Schematic illustration of a dissipative Aharonov-Bohm chain. Each unit cell consists of three sublattice sites $a$, $b$, and $c$, forming a triangular loop threaded by a flux $\phi$. The green, blue and red bounds denote hopping between adjacent sites. See main text for definition of variables.}
	\label{fig:fig1}
\end{figure}

The recent discovery of the non-Hermitian skin effect has stimulated further research activities~\cite{nhtopot4,nhtopot5,murakami,nhse1,nhse2,nhse3,nhse4,nhse5,nhse6,nhsedy1,nhsedy2,nhsedy3}. Under the non-Hermitian skin effect, eigenstates of a system become exponentially localized at boundaries, leading to dramatic changes in the system's band and spectral topology~\cite{nhse3,nhse4}, dynamics~\cite{mastereqeff1,zhushiliang,tianyu,nhsedy1,nhsedy2,nhsedy3}, and spectral symmetry~\cite{longhipt,skinpt,chenpt}. Experimentally, the non-Hermitian skin effect and its consequences have been observed in classical and photonic systems~\cite{teskin,nhtopoe2,classical1,scienceskin}, as well as in a Bose-Einstein condensate of ultracold atoms~\cite{skinatom}.
In the last case, a dissipative Aharonov-Bohm (AB) chain was implemented in the momentum and hyperfine-spin space of the condensate atoms. As illustrated in Fig.~\ref{fig:fig1}, the AB chain consists of a series of triangular AB rings~\cite{yan}, each threaded by a synthetic magnetic flux, realized by engineering the phases of the nearest-neighbor hopping rates. Dissipation is introduced through on-site particle loss for each ring, such that dynamics of atoms that remain in the chain is driven by a non-Hermitian Hamiltonian that features non-trivial band topology. Importantly, the interplay of synthetic flux and dissipation gives rise to a non-reciprocal flow in the bulk that lies at the origin of the non-Hermitian skin effect.
In the experiment, the non-Hermitian skin effect was observed through a directional propagation of atoms along the chain, while the topological edge states were probed through an inverse spectroscopy, where atoms are injected into an empty dissipative AB chain from a bystander state.
Despite its experimental implementation, a systematic study of the topological properties of the dissipative AB chain is missing in the literature.
Further, given the intrinsic difficulty of detecting topological edge states in the presence of non-Hermitian skin effects (as both are localized at the boundary), more variety of detection schemes is desirable.

In this work, we carry out a systematic study of the dissipative AB chain, focusing on its topology and detection. We identify an additional topological phase transition beyond the experimentally demonstrated parameter regime, and derive analytical expressions for the topological transition points using the non-Bloch band theory.
Invoking the theoretical framework of Feshbach projection~\cite{Non1,feshbach}, we derive the transfer rates of the experimentally implemented atom-injection spectroscopy, which are in good agreement with numerical simulations. In the experiment, atoms were injected to an open edge of the AB chain, to detect the topological edge states. Here we show that by injecting atoms into a bulk site far away from the boundary, spectral information under the periodic boundary condition can be obtained from the transfer rate.
We then propose a dynamic detection scheme for the topological edge states.

The paper is organized as follows. In Sec.~II, we review the model Hamiltonian for the dissipative AB chain, and show that it has the non-Hermitian skin effect. In Sec.~III, we characterize its topological properties using the non-Bloch band theory. We then provide a theoretical characterization of the injection spectroscopy in Sec.~IV. In Sec.~V, we discuss the dynamic detection of topological edge states. We summarize in Sec.~VI.

\section{Model}

The non-Hermitian Hamiltonian of the dissipative AB chain illustrated in Fig.~\ref{fig:fig1} is given by~\cite{yan,skinatom}
\begin{equation}
	\begin{aligned}
		H=& \sum_{n=1}^{N}[J_pb^\dagger_na_n+J_sc^\dagger_nb_n+J_se^{i\phi}a^\dagger_nc_n+{H.c.}]\\&+\sum_{n=1}^{N-1} [J_ta^\dagger_{n+1}b_n+{H.c.}]+ \sum_{n=1}^{N}(\Delta-i\gamma)c_n^\dagger c_n.
	\end{aligned}
	\label{eq:Hrealspace}
\end{equation}
Here $a_n(a_n^\dagger)$, $b_n(b_n^\dagger)$ and $c_n(c_n^\dagger)$ are respectively the annihilation (creation) operators for the $a$, $b$ and $c$ sublattice sites of the $n$th unit cell; $J_p$, $J_s$ and $J_t$ are the nearest-neighbour hopping rates; $\Delta$ and $\gamma$ are respectively the on-site potential and the loss rate on site $c$; the phase $\phi\in [0,2\pi)$ corresponds to a synthetic magnetic flux.

\begin{figure}[tbp]
	\includegraphics[width=9cm]{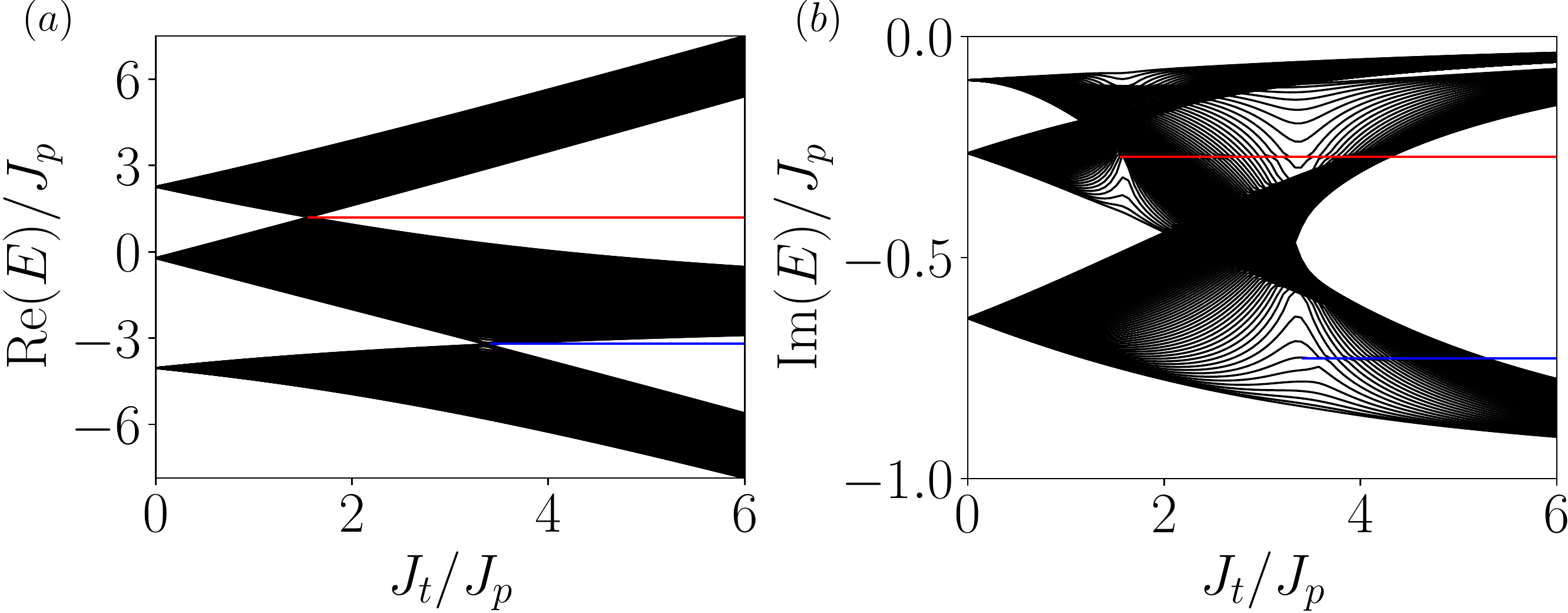}
\caption{Eigenspectra of a dissipative AB chain under the open boundary condition. We take $N=100$ unit cells, $J_s/J_p=2$, $\phi=\pi/2$, $\Delta/J_p=-2$, and $\gamma/J_p=1$ for numerical calculations. (a)(b) The real ($\text{Re}(E)$) and imaginary ($\text{Im}(E)$) components of eigenenergies as functions of $J_t$. The red and blue lines in (a)(b) denote topological edge states, each two-fold degenerate. The two topological transition points are $J_{t,c1}/J_p=1.56$ (associated with edge states in red) and $J_{t,c2}/J_p=3.41$ (associated with those in blue), respectively.}
	\label{fig:fig2}
\end{figure}

Hamiltonian (\ref{eq:Hrealspace}) hosts topological edge states under the open boundary condition.
For finite $\gamma$ and $\phi\notin \{0,\pi\}$, all eigenstates accumulate to the boundaries under the non-Hermitian skin effect. As demonstrated in Ref.~\cite{yan}, this can be understood in the limit $\Delta,\gamma\gg J_s,J_p,J_t$, when Hamiltonian (\ref{eq:Hrealspace}) can be perturbatively reduced to a non-Hermitian Su-Schrieffer-Heeger model with asymmetric hopping. Physically, this is because the interplay of the synthetic flux and on-site loss gives rise to a non-reciprocal flow along the chain. Beyond such a limit, the dissipative AB chain is qualitatively different from a non-Hermitian Su-Schrieffer-Heeger model, particularly for the lack of chiral symmetry. Nevertheless, both the non-Hermitian skin effect and band topology persist as salient features of the dissipative AB chain.

In Fig.~\ref{fig:fig2}, we show typical eigenspectra of the model under the open boundary condition.
Two gap-closing points can be identified, particularly visible in $\text{Re}(E)$, where topological edge states emerge.
The topological transitions are robust under variations of $\phi$ and $\gamma$, their locations however, sensitively depend on these parameters.
In the following, we denote the location of the topological phase transitions as $J_{t,c1}$ and $J_{t,c2}$, which are associated with the edge states in red and blue, respectively, in Fig.~\ref{fig:fig2}. We further denote the corresponding eigenenergies of the topological edge states as $E_{c1}$ and $E_{c2}$, respectively.
Note that the transition at $J_{t,c1}$ was experimentally probed in Ref.~\cite{skinatom}, but not the one at $J_{t,c2}$.

\begin{figure}[tbp]
	\includegraphics[width=8cm]{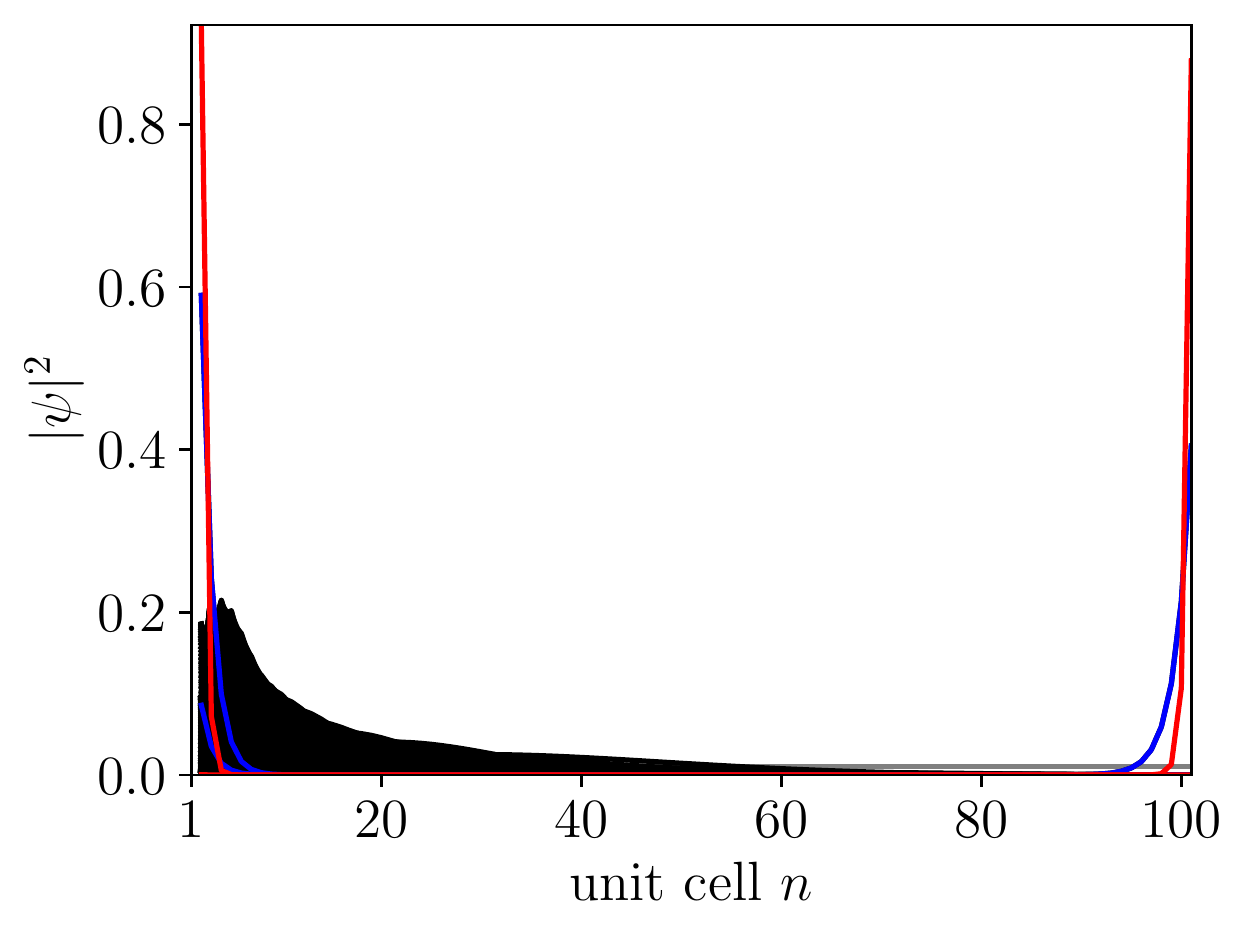}
	\caption{Spacial distribution of the eigenstates for an AB chain with $N=100$. We take $J_t/J_p=5$, while other parameters are the same as those in Fig.~\ref{fig:fig2}. Gray: periodic boundary condition. Black: bulk eigenstates under the open boundary condition. Red and blue: degenerate topological edge states with eigenenergies $E_{c1}$ and $E_{c2}$, respectively.}
	\label{fig:fig3}
\end{figure}

In Fig.~\ref{fig:fig3}, we show the spatial probability distribution of eigen wavefunctions under different boundary conditions. We choose the parameter $J_t/J_p=5$, such that two pairs of topological edge states (indicated by red and blue) exist under the open boundary condition. For finite $\gamma$ and $\phi\notin \{0,\pi\}$, all eigenstates are localized toward the boundaries, indicating the presence of non-Hermitian skin effect. It follows that topological edge states in Fig.~\ref{fig:fig2} can only be accounted for by a non-Bloch topological invariants under the non-Bloch band theory.

\section{Topology under the non-Bloch band theory}

Topological edge states of the dissipative AB chain are characterized by the non-Bloch band theory~\cite{nhtopot4,murakami}. The idea is to take into account the deformation of the bulk eigenstates under the non-Hermitian skin effect, replacing the phase factor $e^{ik}$ of the Bloch waves (under the periodic boundary condition) with a spatial mode factor $\beta(k)=|\beta(k)|e^{ik}$. Here the quasimomentum $k\in [0,2\pi)$, and the trajectory of $\beta(k)$ on the complex plane is known as the generalized Brillouin zone (GBZ), which can be calculated from the Schr\"{o}dinger's equation as shown below.

In the spirit of the non-Bloch band theory, we write the non-Bloch Hamiltonian as
\begin{equation}
	H(\beta)=\left(\begin{array}{ccc}
		0 & J_p+J_t\beta^{-1} & J_se^{i\phi} \\
		J_p+J_t\beta & 0 & J_s \\
		J_se^{-i\phi} & J_s & \Delta-i\gamma
	\end{array}\right).
	\label{eq:hbeta}
\end{equation}
The Schr\"{o}dinger's equation in the GBZ is then $[H(\beta)-E]|\varphi^{R}_j(\beta)\rangle=0$, where
$E$ is the eigenenergy, $|\varphi^{R}_j(\beta)\rangle$ is the right eigenstate, and $j$ is the band index. Sending the determinant of the eigen equation to zero, we have
\begin{equation}
	\begin{aligned}
&\left[J_t\beta+J_p+\frac{e^{-i\phi}J_s^2}{E-(\Delta-i\gamma)}\right]\left[J_t\beta^{-1}+J_p+\frac{e^{i\phi}J_s^2}{E-(\Delta-i\gamma)}\right]\\&-\left[E-\frac{J_s^2}{E-(\Delta-i\gamma)}\right]^2=0.
	\end{aligned}
	\label{eq:betaeq}
\end{equation}

The spatial mode functions $\beta(k)$ can be solved by requiring the two roots of Eq.~(\ref{eq:betaeq}) to have the same magnitude, with $|\beta_1|=|\beta_2|$. We then have~\cite{skinatom}
\begin{equation}
|\beta(k)|=\sqrt{\left|\frac{J_p+\frac{e^{-i\phi}J_s^2}{E-(\Delta-i\gamma)}}{J_p+\frac{e^{i\phi}J_s^2}{E-(\Delta-i\gamma)}}\right|}.
	\label{eq:betamag}
\end{equation}
It is then straightforward to solve for $E$ and $|\beta(k)|$ from Eqs.~(\ref{eq:betaeq}) and (\ref{eq:betamag}) for each $k$. The resulting eigenenergy $E$ gives the eigenspectrum under an open boundary condition.

\begin{figure*}[tbp]
	\includegraphics[width=15cm]{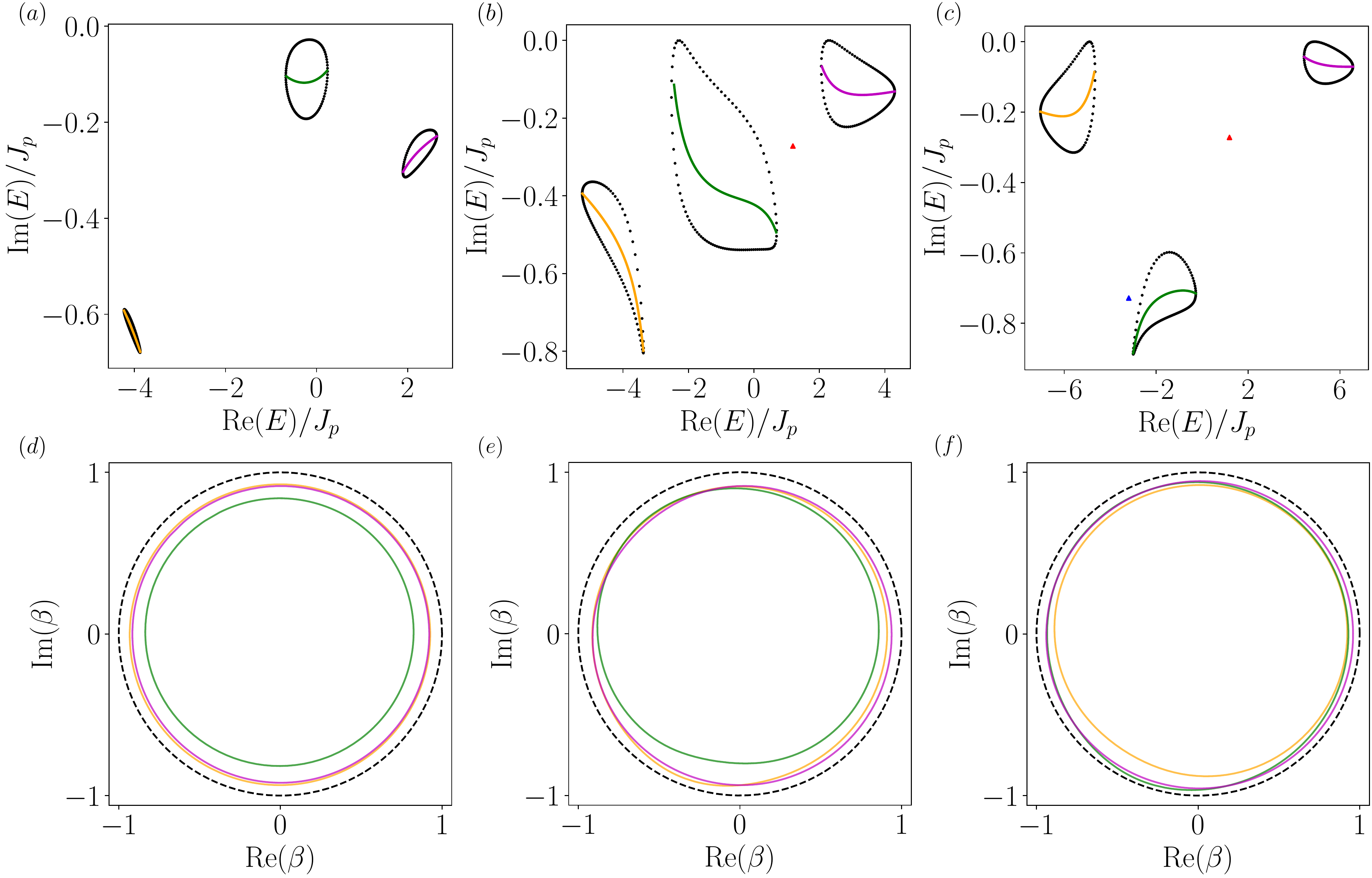}
	\caption{(a)(b)(c) Eigenspectra of Hamiltonian Eq.~(\ref{eq:Hrealspace}) with $N=100$ unit cells on the complex plane. Black: eigenspectra under PBC. Orange, green, and purple: eigenspectra for three different bands under the open boundary condition. Red and blue triangle denote the topological edge states with eigenenergies $E_{c1}$ and $E_{c2}$, respectively.
(d)(e)(f) GBZs on the complex plane. Orange, green, and purple: GBZs for the three different bands in (a)(b)(c). Black dashed line is the unit circle, which corresponds to the conventional Brillouin zone. For (a)(d), $J_t/J_p=0.5$; for (b)(e), $J_t/J_p=2.5$; for (c)(f), $J_t/J_p=5$. Other parameters are the same as those in Fig.~\ref{fig:fig2}.}
	\label{fig:fig4}
\end{figure*}

In Fig.~\ref{fig:fig4}(a)(b)(c), we show the eigenspectra for different parameters, under both the periodic (dots) and the open boundary conditions (solid curves). Under the periodic boundary condition, the eigenenergies of each band form a closed spectra loop, consistent with the well-known spectral topology of the non-Hermitian skin effect. By contrast, under the open boundary condition, the eigenenergies collapse to open arcs within the closed loops. In Fig.~\ref{fig:fig4}(b)(c), the discrete red and blue dots outside the spectra loops correspond to the topological edge states in Fig.~\ref{fig:fig2}.

\begin{figure}[tbp]
	\includegraphics[width=8cm]{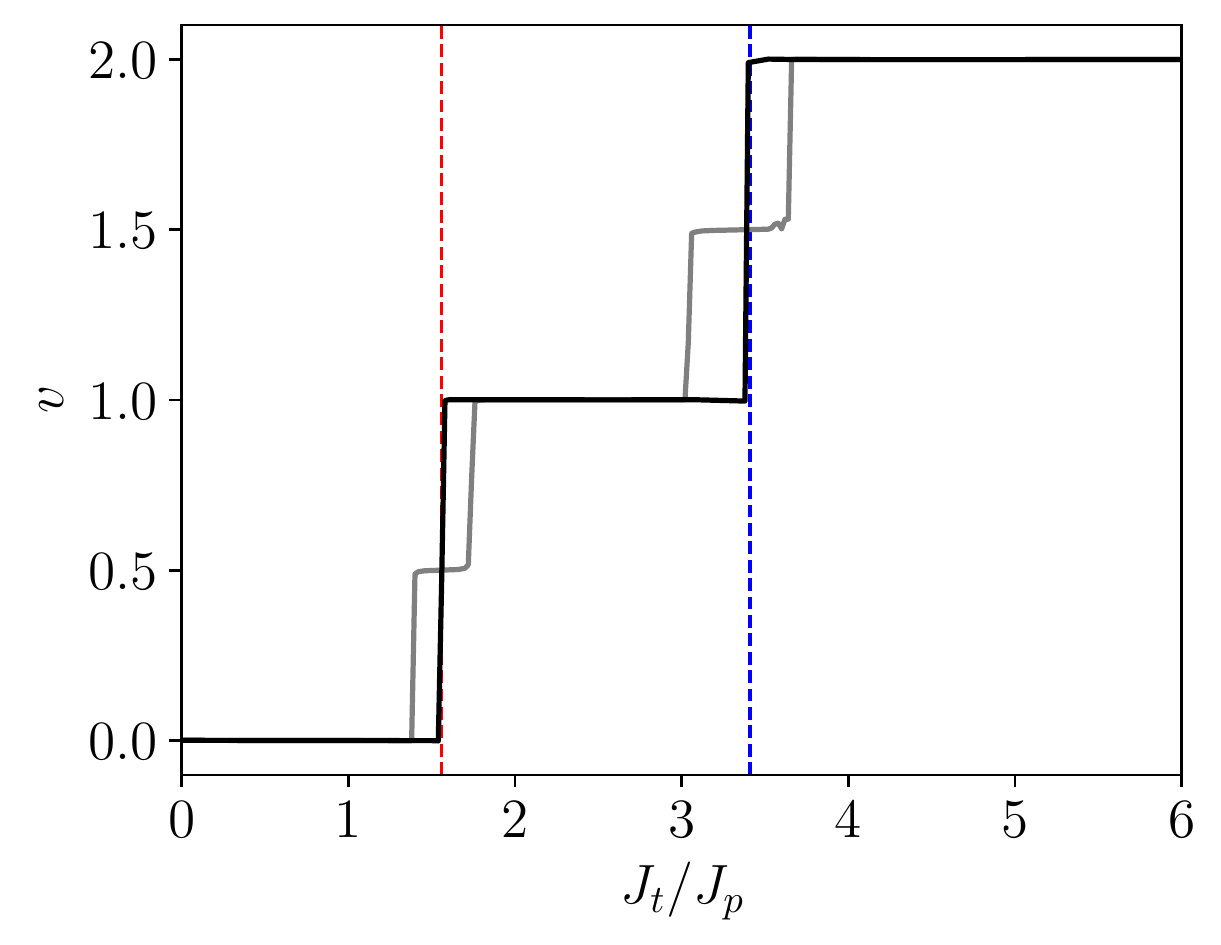}
	\caption{Bloch (gray) and non-Bloch (black) winding numbers. The vertical dashed lines in red and blue denote $J_{t,c1}$ and $J_{t,c2}$, respectively. All parameters are the same as those in Fig.~\ref{fig:fig2}.}
	\label{fig:fig5}
\end{figure}

In Fig.~\ref{fig:fig4}(d)(e)(f), we plot the GBZs of the three bands under the parameters of Fig.~\ref{fig:fig4}(a)(b)(c), respectively. For all cases, the calculated $|\beta(k)|<1$, and the GBZs are within the unit circle. This indicates that under the open boundary condition, all eigenstates accumulate to the left boundary (toward small unit-cell index $n$).

\begin{figure*}[tbp]
	\includegraphics[width=15cm]{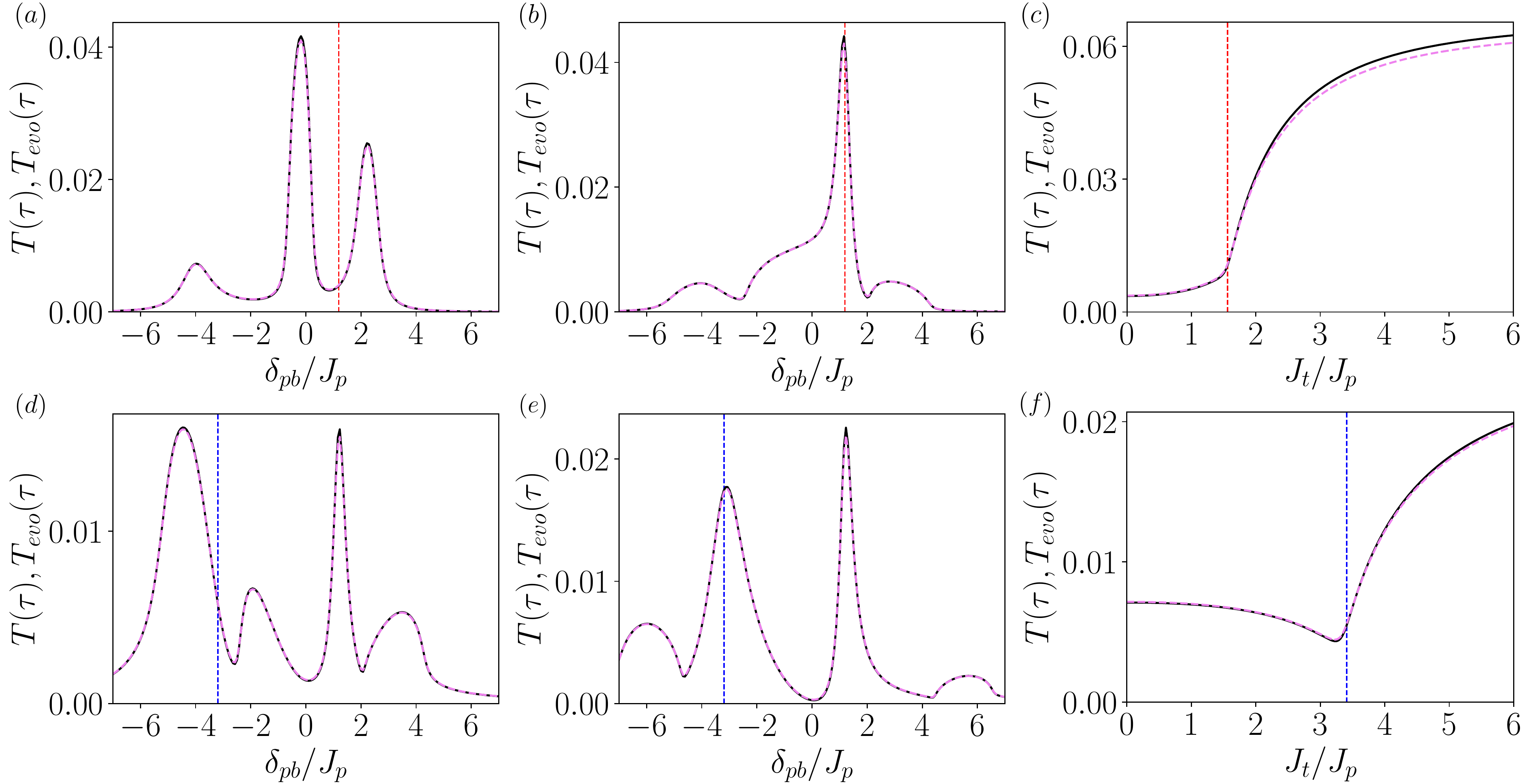}
	\caption{Transfer rate for the inverse spectroscopy, for $|f\rangle$ located at an open boundary. (a)(b)(c) The probe Hamiltonian couples the bystander state to the $b$ sublattice site of the $N$th unit cell (the right-most unit cell on the edge), with (a) $J_t/J_p=0.5$, (b) $J_t/J_p=2.5$, and (c) $\delta_{pb}/J_p=\mathrm{Re}(E_{c1})$. Here $E_{c1}$ is the energy of the topological edge state. The dashed vertical lines in (a)(b) correspond to $\mathrm{Re}(E_{c1})$, and the dashed line in (c) corresponds to $J_{t,c1}$.
(d)(e)(f)  The probe Hamiltonian couples the bystander state to the $c$ sublattice site of the $N$th unit cell (the right-most unit cell on the edge), with (d) $J_t/J_p=2.5$, (e) $J_t/J_p=5$, and (f) $\delta_{pb}/J_p=\mathrm{Re}(E_{c2})$.
The dashed vertical lines in (d)(e) correspond to $\mathrm{Re}(E_{c2})$, and the dashed line in (f) corresponds to $J_{t,c2}$.
For all subplots, the black solid lines and the magenta dashed lines are respectively the theoretically predicted transfer rate using Eq.~(\ref{eq:tratio}), and the numerically simulated transfer rates from Eq.~(\ref{eq:tratioevo}). For all figures, $J_{pb}/J_p=0.01$, $\tau J_p=40\pi$. Other parameters are the same as those in Fig.~\ref{fig:fig2}.}
	\label{fig:fig6}
\end{figure*}

\begin{figure}[tbp]
	\includegraphics[width=9cm]{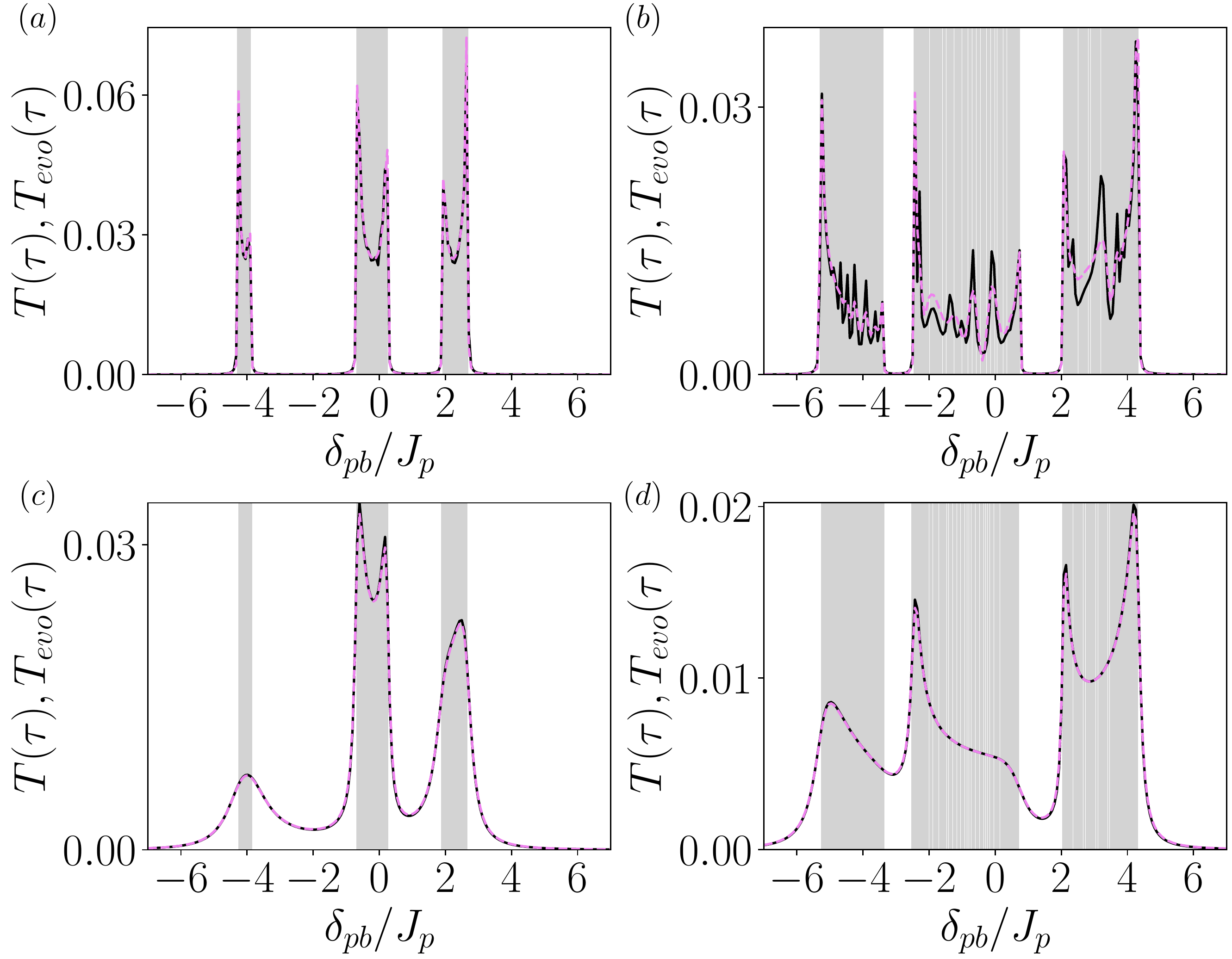}
	\caption{Transfer rate for the inverse spectroscopy, as the probe Hamiltonian is coupled to the $b$ sublattice site of the $51$st unit cell, which is deep in the bulk.
(a)(b) Hermitian case with $\gamma=0$. (c)(d) Non-Hermitian case with $\gamma/J_p=1$. We also take $J_t/J_p=0.5$ in (a)(c), and $J_t/J_p=2.5$ in (b)(d), while other parameters are the same as those in Fig.~\ref{fig:fig6}. For all subplots, {$\tau J_p=40\pi$}, the black solid lines and the magenta dashed lines are respectively the theoretically predicted transfer rate using Eq.~(\ref{eq:tratio}), and the numerically simulated transfer rates from Eq.~(\ref{eq:tratioevo}). The shaded regions in gray indicate the real components of the spectra under the
periodic boundary condition.
}
\label{fig:fig7}
\end{figure}

\begin{figure}[tbp]
	\includegraphics[width=9cm]{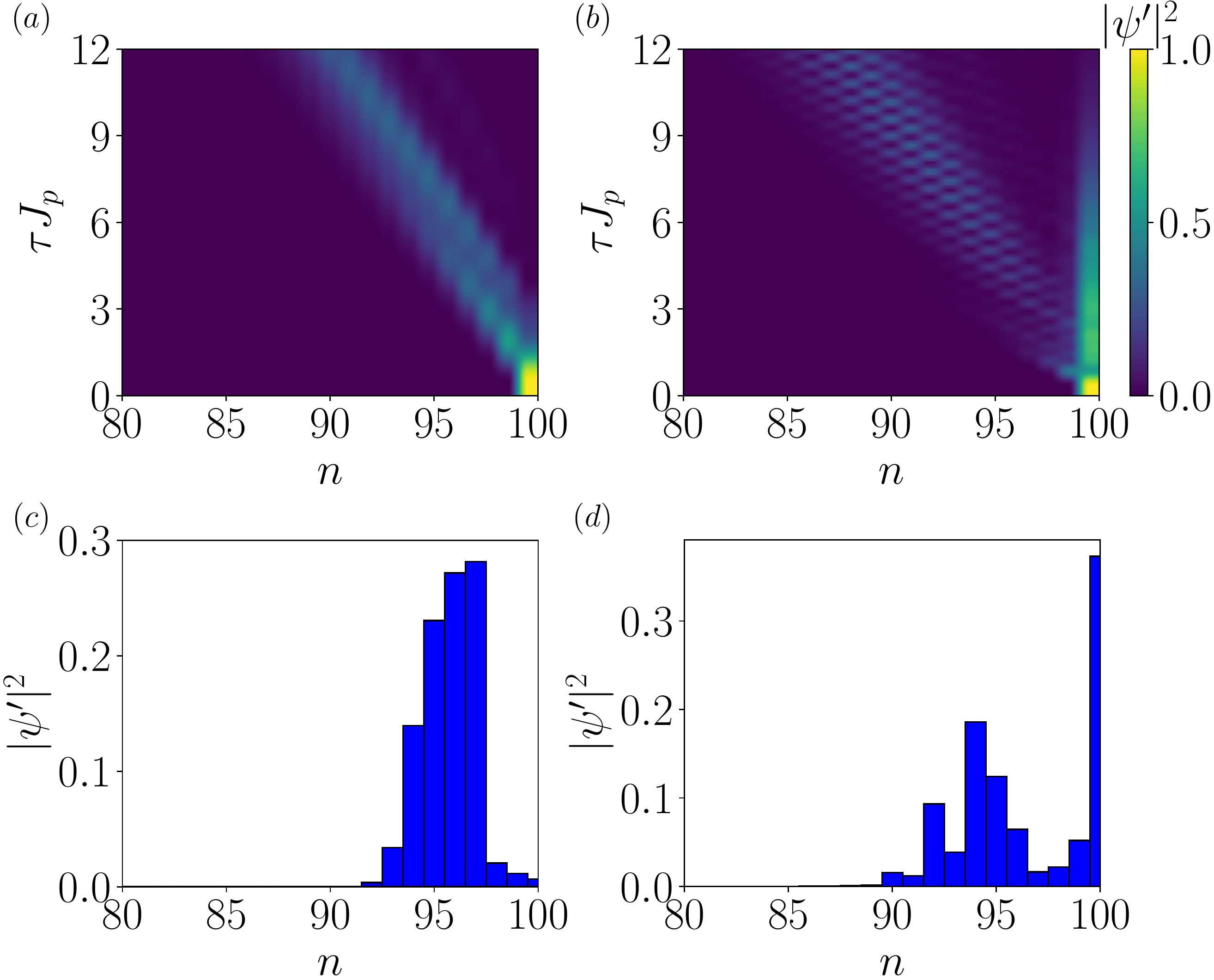}
	\caption{Boundary dynamics for the detection of topological edge states.
(a)(b) Color contour for the normalized occupation distribution as a function of time. (b)(d) The normalized occupation distribution at the time $\tau J_p=6$. We take $J_t/J_p=1$ in (a)(c), and $J_t/J_p=5$ in (b)(d). Other parameters are the same as those in Fig.~\ref{fig:fig2}.}
	\label{fig:fig8}
\end{figure}

We are now in a position to calculate the non-Bloch winding number, which can restore the bulk-boundary correspondence and predict the existence and number of topological edge states. Unlike the non-Hermitian Su-Schrieffer-Heeger model, the dissipative AB chain features three bands.
The non-Bloch winding number $\nu$ is {defined through the global Berry phase, which is the sum of the Berry phases} of all three bands, with
\begin{equation}
	\nu=\frac{1}{2\pi}\sum_j\Theta_{j}.
	\label{eq:windnum}
\end{equation}
Here the {Berry phase} of the $j$th band is given by
{\begin{equation}
		\Theta_{j}=i \oint_{\mathrm{GBZ}_j} \mathrm{d}\beta\left\langle \varphi^L_{j}(\beta)\left|\partial_{\beta}\right| \varphi^R_{j}(\beta)\right\rangle,
		\label{eq:berryphase}
	\end{equation}}
where the right and left eigenstates of $H(\beta)$ are defined as $H(\beta)|\varphi_j^R(\beta)\rangle=E_j|\varphi_j^R(\beta)\rangle$ and  $H^\dagger(\beta)|\varphi_j^L(\beta)\rangle=E_j^*|\varphi_j^L(\beta)\rangle$, respectively.
The integration in Eq.~(\ref{eq:berryphase}) is over { the GBZ of the $j$th band.}
When $\beta$ is replaced by $e^{ik}$ in Eq.~(\ref{eq:berryphase}), the non-Bloch winding number is reduced to the Bloch winding number which characterizes the band topology of the system under the periodic boundary condition.

In Fig.~\ref{fig:fig5}, we show the Bloch (gray) and non-Bloch (black) winding numbers. The non-Bloch winding number is quantized to integers, and changes its value at topological transitions that are consistent with the gap-closing points in Fig.~\ref{fig:fig2}. By contrast, the Bloch winding number can take half-integer values, and it does not indicate the topological transitions of the system under the open boundary condition. Note that half-integer winding numbers have previously been reported in the non-Hermitian, asymmetric Su-Schrieffer-Heeger model~\cite{nhse5,halfwinding}, where explicit geometric interpretations can be found based on its chiral symmetry. While chiral symmetry is absent in our model, the origin of these half-integer winding numbers, and their general relation to the non-Hermitian skin effect, are interesting open questions.

The topological transition points can be analytically determined from the gap-closing condition.
At the gap-closing pint, GBZs of two different bands intersect on the complex plane at the same eigenenergies.
It follows that Eq.~(\ref{eq:betaeq}) features a double root at the topological transition. This is satisfied for
\begin{equation}
	\begin{aligned}
		J_{t,cj}=&\sqrt{\left|(J_p+\frac{e^{i\phi}J_s^2}{E_{cj}-(\Delta-i\gamma)})\cdot
		(J_p+\frac{e^{-i\phi}J_s^2}{E_{cj}-(\Delta-i\gamma)})\right|},
	\end{aligned}
	\label{eq:jtc}
\end{equation}
with $j=1,2$. And the roots are given by
\begin{align}
	E_{c1}=\frac{(\Delta-i\gamma)+\sqrt{(\Delta-i\gamma)^2+4J_s^2}}{2},
	\label{eq:ec1}\\
    E_{c2}=\frac{(\Delta-i\gamma)-\sqrt{(\Delta-i\gamma)^2+4J_s^2}}{2}, \label{eq:ec2}
\end{align}
which correspond to the energies of the topological edge states emerging at the two transition points in Fig.~\ref{fig:fig1}.
Both $J_{t,cj}$ and $E_{cj}$ calculated from Eqs.~(\ref{eq:jtc})(\ref{eq:ec1})(\ref{eq:ec2}) are in excellent agreement with the numerically calculated eigenspectra in Fig.~\ref{fig:fig2}. Note that we take the positive branch for the square roots in Eqs.~(\ref{eq:ec1})(\ref{eq:ec2}).

To close this section, we discuss the symmetry of Hamiltonian (\ref{eq:hbeta}).
While it does not have chiral symmetry, Hamiltonian (\ref{eq:hbeta}) is symmetric under the following transformation:
$\Gamma H^{T}(\beta) \Gamma^{-1}=H(\beta)$, where
$\Gamma=\left(\begin{array}{ccc}0&1&0\\1&0&0\\0&0&e^{-i\phi} \end{array}\right)$. In the Hermitian limit with $\gamma=0$, the symmetry is reduced to $\Gamma H^{T}(k) \Gamma^{-1}=H(k)$, where $k$ is then the quasi-momentum in the conventional Brillouin zone. We have checked that such a symmetry protects the two-fold degeneracy of the topological edge states emerging from either phase transitions. {Note that, while the Berry phases $\Theta_{j}$ are quantized to multiples of $\pi$ in the presence of such a symmetry, they are no longer so when the symmetry is broken. By contrast, the non-Bloch winding number is always quantized, since the global Berry phase, when integrated over the GBZ, is always integer multiples of $2\pi$.}

In the real lattice space, the symmetry operation can be further decomposed into $\Gamma=PC_+$, where $P: a_n \rightarrow a_{N-n}, \,\, b_n \rightarrow b_{N-n}, \,\, c_n \rightarrow c_{N-n}$; $C_+: a_n \rightarrow b_{N-n}, \,\, b_n \rightarrow a_{N-n}, \,\, c_n \rightarrow e^{-i\phi}c_{N-n}$.

We identify $P$ and $C_+$ as the inversion and the non-Hermitian variant of the time-reversal operators, respectively. In particular, $C_+$ can be identified with the $\text{TRS}^\dag$ symmetry in Ref.~\cite{nhtopot2}. Physically, the combined inversion and time-reversal symmetry is understood from the observation that the dissipative AB chain remains invariant by simultaneously reversing the flux and the lattice, but not either alone.

\section{Detecting topological edge states and band structure}

In the experiment~\cite{skinatom}, a momentum-resolved Bragg spectroscopy was applied to detect the topological edge states, where atoms are injected into an edge site of the AB chain. In this section, we provide a theoretical description for the atom-injection spectroscopy, and show that a similar detection scheme can be applied to probe the band structure. We then propose an alternative dynamic detection scheme for the topological edge states.

\subsection{Injection spectroscopy}
We consider coupling atoms in a bystander state $|d\rangle$ to a local site $|f\rangle$ of the dissipative AB chain. Site $|f\rangle$ can be any one of the sublattice sites $|a\rangle$, $|b\rangle$ or $|c\rangle$. It can be on the edge, as is the case in the experiment, or in the bulk, {far away} from any boundaries. The chain is originally empty, such that the scheme is similar in spirit to the inverse radio-frequency spectroscopy. The probe Hamiltonian reads
\begin{equation}
	H_{pb}=J_{pb}d^\dagger f+J_{pb}f^\dagger d+\delta_{pb} d^\dagger d.
	\label{eq:Hpb}
\end{equation}
Here, $d$ ($d^\dagger$), $f$ ($f^\dagger$) respectively denote the annihilation (creation) operators for state $|d\rangle$ and $|f\rangle$. $J_{pb}$ is the coupling rate between $|d\rangle$ and $|f\rangle$, $\delta_{pb}$ is the detuning of the coupling frequency with respect to the transition $|d\rangle\rightarrow |f\rangle$. {Here the overall dynamics is governed by the Hamiltonian $H^\prime=H+H_{pb}$.}

Following the practice of Feshbach projection~\cite{feshbach,feshbach2,cohen}, we define the projection operators $P=|d\rangle\langle d|$ and $Q=\mathbf{I}-P$. The effective Hamiltonian in the subspace of the bystander state $|d\rangle$ is then
\begin{equation}
	H_{\mathrm{eff}}(E) =H^\prime_{P P}+H^\prime_{P Q} \frac{1}{E-H^\prime_{Q Q}} H^\prime_{Q P},
	\label{eq:Heffform}
\end{equation}
where
\begin{equation}
	\begin{aligned}
		H^\prime_{P P} =\delta_{pb}d^\dagger d,&\quad H^\prime_{P Q} =J_{pb}d^\dagger f, \\
		H^\prime_{Q P} =J_{pb}f^\dagger d,&\quad H^\prime_{Q Q} =H.
	\end{aligned}
	\label{eq:hproj}
\end{equation}

It can be shown straightforwardly that
\begin{equation} H_{\mathrm{eff}}(E)=\left(\delta_{pb}+\sum_j\frac{J_{pb}^2\psi_{j,f}^R\psi_{j,f}^{L*}}{E-E_j}\right)d^\dagger d,
\label{eq:heffcal1}
\end{equation}
where $\psi_{j,f}^{R}=\langle f|\psi_j^R\rangle$ and $\psi_{j,f}^{L*}=\langle\psi_j^L|f\rangle$, and the right and left eigenstates of $H$ are defined as
$H|\psi_j^R\rangle=E_j|\psi_j^R\rangle, H^\dagger|\psi_j^L\rangle=E_j^*|\psi_j^L\rangle$.
The effctive Hamiltonian Eq.~(\ref{eq:heffcal1}) is dissipative, where the dissipation is due to the atom transfer from state $|d\rangle$ to the AB chain. We define the transfer rate
\begin{equation}
	T(\tau)=1-\exp[-2R(\delta_{pb})\tau],
	\label{eq:tratio}
\end{equation}
which describes the probability for an atom to be transferred from $|d\rangle$ to the chain within the evolution time $\tau$. Here
\begin{equation}
	R(\delta_{pb})=-\operatorname{Im}\left(\langle d|H_{\mathrm{eff}}(E)|d\rangle\right)=-\operatorname{Im} \sum_j\frac{J_{pb}^2\psi_{j,f}^R\psi_{j,f}^{L*}}{\delta_{pb}-E_j+i0^{+}}.
	\label{eq:trate}
\end{equation}
The term $i0^+$ in the denominator ensures that Eq.~(\ref{eq:trate}) recovers the familiar form of the Fermi's golden rule in the Hermitian case. Note that the same Eq.~(\ref{eq:trate}) can be derived using the linear response theory (see Appendix).

The transfer rate can also be evaluated through numerical simulation of the system dynamics. Initialized in the state $|d\rangle$ at the initial time $t=0$, the time evolved state at time $\tau$ is then
$|\psi(\tau)\rangle=e^{-iH^\prime\tau}|d\rangle$. {Note that the non-normalized nature of $|\psi(\tau)\rangle$ corresponds to loss of atoms from the dissipative AB chain.}
The transfer rate can be expressed as
\begin{equation}
	T_{evo}(\tau)=1- |\langle d|e^{-iH^\prime \tau}|d\rangle|^2.
	\label{eq:tratioevo}
\end{equation}
Due to the perturbative nature of Eq.~(\ref{eq:tratio}), we expect that the transfer rates calculated from Eq.~(\ref{eq:tratio}) and Eq.~(\ref{eq:tratioevo}) are very close to each other, provided that $J_{pb}/J_p\ll 1$ and $J_{pb}^2\tau\ll 1\ll J_p\tau$.  This is indeed the case as we show below.

For the detection of the topological edge states, we follow the practice of Ref.~\cite{skinatom}, and consider $|f\rangle$ to be on the edge of the AB chain. The calculated transfer rates are plotted in Fig.~\ref{fig:fig6}. While results from Eq.~(\ref{eq:tratio}) and Eq.~(\ref{eq:tratioevo}) agree well with one another, signals of the edge states are visible near the appropriate detuning $\delta_{pb}$. Specifically, in Fig.~\ref{fig:fig6}(a)(b), we aim to detect the edge state with energy $E_{c1}$. The state has a large support on sublattice site $b$, we therefore set $|f\rangle$ on site $b$ of the right-most unit cell on the edge. When $J_{t}<J_{t,c1}$, $T(\tau)$ exhibits a valley at $\delta_{pb}=\mathrm{Re}(E_{c1})$ [Fig.~\ref{fig:fig6}(a)], indicating the presence of a band gap. When $J_{t}>J_{t,c1}$, a peak emerges at $\delta_{pb}=\mathrm{Re}(E_{c1})$ [Fig.~\ref{fig:fig6}(b)], suggesting the presence of an in-gap edge state.
In Fig.~\ref{fig:fig6}(c), we show the transfer rate at $\delta_{pb}=\mathrm{Re}(E_{c1})$ as a function
of $J_t$, which clearly indicates a phase transition near $J_{t,c1}$.

Similarly, in Fig.~\ref{fig:fig6}(d)(e), we aim to probe the edge state with energy $E_{c2}$.
Since the edge states now have a large support on sublattice site $c$, we set $|f\rangle$ on site $c$ of the right-most unit cell.
We find that a peak appears near $\delta_{pb}=\mathrm{Re}(E_{c2})$ in the transfer rate when $J_{t}>J_{t,c2}$ [Fig.~\ref{fig:fig6}(d)(e)], consistent with the emergence of the topological edge states. Likewise,
the topological phase transition is clearly visible near $J_t=J_{t,c2}$ in Fig.~\ref{fig:fig6}(f).

Compared to the Bragg spectroscopy implemented in Ref.~\cite{skinatom}, for the numerical simulations here, we consider a weaker probe ($J_{\text{pb}}\sim h\times 12$ Hz, $h$ being the planck constant), and a longer probe time ($\tau\sim 16$ ms). Such optimization leads to a more faithful detection with a better resolution.

Alternatively, we can couple the bystander state to a sublattice site in the bulk to reveal the global spectral features under the periodic boundary condition. The results are plotted in Fig.~\ref{fig:fig7}. In the Hermitian case with $\gamma=0$ [Fig.~\ref{fig:fig7}(a)(b)], the transfer rate shows sharp edges at the band edge, revealing both the band continuum and the band gaps. For the dissipative AB chain with finite $\gamma$ [Fig.~\ref{fig:fig7}(c)(d)], the transfer-rate profiles are broadened due to the imaginary components of the eigenspectra. Nevertheless, the band gaps are still visible as valleys in the profile.
In relation to the experiment in Ref.~\cite{skinatom}, injecting atoms into the bulk offers a complementary detection scheme for the topological phase transition, by observing the closing of the band gaps.

\subsection{Dynamic detection of edge states}

Topological edge states can also be detected through dynamics close to the boundary.
Under the non-Hermitian skin effect, eigenstates of the AB chain accumulate to one of the edges.
The idea is to initialize the state near the opposite edge, and observe the time-dependent population along the chain. While the non-Hermitian skin effect would drive the population toward the other edge, topological edge states should remain near the initial site. To quantitatively characterize the phenomena, we define the normalized occupation
\begin{equation}
	|\psi^\prime(\tau,n)|^2=\sum_{j=a,b,c}\frac{|\langle n,j|\psi(\tau)\rangle|^2}{\langle\psi(\tau)|\psi(\tau)\rangle},
	\label{eq:norocu}
\end{equation}
which indicates the spatial distribution of the state at the time $\tau$.

In Fig.~\ref{fig:fig8}, we show the time evolution of the probability in the topological trivial [Fig.~\ref{fig:fig8}(a)(c)] and non-trivial regions [Fig.~\ref{fig:fig8}(b)(d)].
In the topological trivial region, the time-evolved state diffuses into the bulk without much occupation at the boundary. By contrast, in the topological non-trivial region, the time-evolved state still exhibits
a peak at the boundary, together with the diffusive dynamics into the bulk.
Note that the normalized occupation at the boundary decreases with time because there are bulk eigenstates that decay slower than the edge states. The detection scheme therefore should only work at intermediate times. Nevertheless, such a dynamic detection is readily accessible in experiments, and provides a direct signal of the non-Hermitian skin effect.

\section{Conclusion}
In conclusion, we have characterized the topological features of a dissipative AB chain in detail, and provided a theoretical description for the recently implemented atom-injection spectroscopy. We show that the injection spectroscopy can be applied to the bulk sites and resolve the band structure of the system under a periodical boundary condition. We further propose an alternative detection scheme for the topological edge states. Our studies are helpful for future experimental study of the dissipative AB chain in relevant quantum simulation systems.

\begin{acknowledgments}
This work has been supported by the Natural Science Foundation of China (Grant Nos. 11974331) and the National Key R\&D Program (Grant Nos. 2016YFA0301700, 2017YFA0304100).
\end{acknowledgments}

\appendix

\renewcommand{\thesection}{\Alph{section}}
\renewcommand{\thefigure}{A\arabic{figure}}
\renewcommand{\thetable}{A\Roman{table}}
\setcounter{figure}{0}
\renewcommand{\theequation}{A\arabic{equation}}
\setcounter{equation}{0}

\section*{Appendix}

In this Appendix, we show that the expression in Eq.~(\ref{eq:trate}) can also be derived from the linear response theory~\cite{linresponse1,linresponse2}.
Given the probe Hamiltonian Eq.~(\ref{eq:Hpb}), we consider the correlation function
\begin{equation}
	D(t,t^\prime)=-i\theta(t-t^\prime)\left\langle\left[f^\dagger(t)d(t),d^\dagger(t^\prime)f(t^\prime)\right]\right\rangle.\
	\label{eq:retarded_Gf}
\end{equation}
In the frequency space, the correlation function becomes
\begin{equation}
	D(i\delta_{pb})=\frac{1}{\beta}\sum_j\sum_nG_{d}(i\omega_n-i\delta_{pb})G_{f}(E_j,i\omega_n),
	\label{eq:Matsu_func}
\end{equation}
where $G_d(i\omega_n)=1/i\omega_n$ is the Green's function for state $|d\rangle$, and $G_f$ is the Green's functions for state $|f\rangle$, with
\begin{equation}
 G_{f}(E_j,i\omega_n)=\frac{\langle 0|f|\psi_j\rangle\langle \psi_j|f^\dagger|0\rangle}{i\omega_n-E_j}=\frac{\psi_{j,f}^R\psi_{j,f}^{L*}}{i\omega_n-E_j}.
 \label{eq:Gx}
\end{equation}
Here $\omega_n$ are the bosonic Matsubara frequencies, $E_j$ and $|\psi_j\rangle $ are respectively the eigenenergy and eigenstate of the bulk Hamiltonian Eq.~(\ref{eq:Hrealspace}). The initial state $|0\rangle$ corresponds to an empty lattice, particularly with no occupation on site $f$. After analytic continuation, we derive the response function $R(\delta_{pb})$ as
\begin{align}
	R(\delta_{pb})&=-J_{pb}^2\operatorname{Im} D\left(i \delta_{pb} \rightarrow \delta_{pb}+i 0^+\right)\nonumber\\
&=-\operatorname{Im} \sum_j\frac{J_{pb}^2\psi_{j,f}^R\psi_{j,f}^{L*}}{\delta_{pb}-E_j+i0^{+}}.
\end{align}
We thus reproduces Eq.~(\ref{eq:trate}) by applying the linear response theory to our non-Hermitian system.


\begin{thebibliography}{99}

\bibitem{photonics1} T. Ozawa, H. M. Price, A. Amo, N. Goldman, M. Hafezi, L. Lu, M. C. Rechtsman, D. Schuster, J. Simon, O. Zilberberg, and I. Carusotto, Rev. Mod. Phys. {\bf 91}, 015006 (2019).

\bibitem{photonics3} {\c{S}}. K. {\"O}zdemir, S. Rotter, F. Nori, and L. Yang, Nat. Mater. \textbf{18}, 783 (2019).
\bibitem{photonics4} M. A. Miri and A. Al\'u, Science \textbf{363}, eaar7709 (2019).
\bibitem{luole}
J. Li, A. K. Harter, J. Liu, L. de Melo, Y. N. Joglekar, and L. Luo, Nat. Commun. \textbf{10}, 855 (2019).
\bibitem{bryce} S. Lapp, J. Ang'ong'a, F. A. An, and B. Gadway, New J. Phys. \textbf{21}, 045006 (2019).
\bibitem{yan} W. Gou, T. Chen, D. Xie, T. Xiao, T.-S. Deng, B. Gadway, W. Yi, and B. Yan, Phys. Rev. Lett. \textbf{124}, 070402 (2020).
\bibitem{NHSOCexp} Z. Ren, D. Liu, E. Zhao, C. He, K. K. Pak, J. Li, and G. Jo, Nat. Phys. {\bf 18}, 385-389 (2022).


\bibitem{trappedion3}P. Schindler, M. M\"{u}ller, D. Nigg, J. T. Barreiro, E. A. Martinez, M. Hennrich, T. Monz, S. Diehl, P. Zoller and R. Blatt, Nat. Phys. \textbf{9}, 361-367 (2013).
\bibitem{chenion} W.-C. Wang, Y.-L. Zhou, H.-L. Zhang, J. Zhang, M.-C. Zhang, Y. Xie, C.-W. Wu, T. Chen, B.-Q. Ou, W. Wu, H. Jing, and P.-X. Chen, Phys. Rev. A \textbf{103}, L020201 (2021).
\bibitem{trappedion4}L. Ding, K. Shi, Q. Zhang, D. Shen, X. Zhang, and W. Zhang, Phys. Rev. Lett. \textbf{126}, 083604 (2021).
\bibitem{Non1} N. Moiseyev, Non-Hermitian quantum mechanics, {\it Cambridge University Press} (2011).
\bibitem{Uedareview} Y. Ashida, Z. Gong, and M. Ueda, Adv. Phys. \textbf{69}, 3 (2020).
\bibitem{molmer}J. Dalibard, Y. Castin, and K. M\o lmer, Phys. Rev. Lett. {\bf68}, 580 (1992).
\bibitem{michael}H. J. Carmichael, Phys. Rev. Lett. {\bf70}, 2273 (1993).
\bibitem{weimer} H. Weimer, A. Kshetrimayum, and R. Or{\'u}s, Rev. Mod. Phys. {\bf93}, 015008 (2021).

\bibitem{mastereqeff1}F. Song, S. Yao, and Z. Wang,	Phys. Rev. Lett. {\bf 123}, 170401 (2019).
	\bibitem{mastereqeff2}N, Shibata and H, Katsura, Phys. Rev. B {\bf 99}, 174303 (2019).
\bibitem{zhushiliang} P. He, Y.-G. Liu, J.-T. Wang, and S.-L. Zhu, Phys. Rev. A {\bf 105}, 023311 (2022).
\bibitem{tianyu} T. Li, Y.-S. Zhang, and W. Yi, Phys. Rev. B {\bf 105}, 125111 (2022).
\bibitem{PT1}
C. M. Bender and S. Boettcher, Phys. Rev. Lett. \textbf{80}, 5243 (1998).
\bibitem{photonics2} R. El-Ganainy, K. Makris, M. Khajavikhan, Z. H. Musslimani, S. Rotter, and D. N. Christodoulides, Nat. Phys. \textbf{14}, 11-19 (2018).
\bibitem{sensor1}J. Wiersig, Phys. Rev. Lett. {\bf 112}, 203901 (2014).
\bibitem{sensor2}H. Hodaei, A. U. Hassan, S. Wittek, H. Garcia-Gracia, R.
	El-Ganainy, D. N. Christodoulides, Nature (London) {\bf 548}, 187 (2017).
\bibitem{sensor3}W. Chen, S. K. \"{O}zdemir, G. Zhao, J. Wiersig, and L. Yang, Nature (London) {\bf 548}, 192 (2017).
	\bibitem{nhtopot1}Z. Gong, Y. Ashida, K. Kawabata, K. Takasan, S. Higashikawa, and M. Ueda, Phys. Rev. X {\bf 8}, 031079 (2018).
	\bibitem{nhtopot2}K. Kawabata, K. Shiozaki, M. Ueda, and M. Sato, Phys. Rev. X {\bf 9}, 041015 (2019).
	\bibitem{nhtopot3}D. Leykam, K. Y. Bliokh, C. Huang, Y. D. Chong, and F. Nori.	Phys. Rev. Lett. {\bf 118}, 040401 (2017).
	\bibitem{nhtopoe1}J. M. Zeuner, M. C. Rechtsman, Y. Plotnik, Y. Lumer, S. Nolte, M. S. Rudner, M. Segev, and A. Szameit, Phys. Rev. Lett. {\bf 115}, 040402 (2015).
\bibitem{nhtopoe15} X. Zhan, L. Xiao, Z. Bian, K. Wang, X. Qiu, B. C. Sanders, W. Yi, and P. Xue, Phys. Rev. Lett. {\bf 119}, 130501 (2017).
\bibitem{nhtopoe16} L. Xiao, X. Zhan, Z. Bian, K. K. Wang, X. Zhang, X. P. Wang, J. Li, K. Mochizuki, D. Kim, N. kawakami, W. Yi, H. Obuse, B. C. Sanders, and P. Xue, Nat. Phys. {\bf 13}, 1117 (2017).


	\bibitem{nhtopot4}S. Yao and Z. Wang, Phys. Rev. Lett. {\bf 121}, 086803 (2018).
	\bibitem{nhtopot5}S. Yao, F. Song, and Z. Wang, Phys. Rev. Lett. {\bf 121}, 136802 (2018).
\bibitem{murakami}K. Yokomizo and S. Murakami, Phys. Rev. Lett. {\bf 123}, 066404 (2019).
	\bibitem{nhse1}C. H. Lee and R. Thomale, Phys. Rev. B \textbf{99}, 201103(R) (2019).
	\bibitem{nhse2}A. McDonald, T. Pereg-Barnea, and A. A. Clerk, Phys. Rev. X \textbf{8}, 041031 (2018).
	\bibitem{nhse3}K. Zhang, Z. Yang, and C. Fang, Phys. Rev. Lett. \textbf{125}, 126402 (2020).
	\bibitem{nhse4}N. Okuma, K. Kawabata, K. Shiozaki, and M. Sato, Phys. Rev. Lett. \textbf{124}, 086801 (2020).
	\bibitem{nhse5}T.-S. Deng and W. Yi, Phys. Rev. B \textbf{100}, 035102 (2019).
	\bibitem{nhse6}Z. Yang, K. Zhang, C. Fang, and J. Hu, Phys. Rev. Lett. \textbf{125}, 226402 (2020).
	\bibitem{nhsedy1}S. Longhi, Phys. Rev. Research \textbf{1}, 023013 (2019).
	\bibitem{nhsedy2}T. Li, J.-Z. Sun, Y.-S. Zhang, and W. Yi, Phys. Rev.
	Research \textbf{3}, 023022 (2021).
	\bibitem{nhsedy3}S. Longhi, Phys. Rev. B \textbf{102}, 201103(R) (2020).
\bibitem{longhipt} S. Longhi, Opt. Lett. {\bf 44}, 5804-5807 (2019).
\bibitem{skinpt} L. Xiao, T. Deng, K. Wang, Z. Wang, W. Yi, and P. Xue, Phys. Rev. Lett. {\bf 126}, 230402 (2021).
\bibitem{chenpt} Y. Liu, Q. Zhou, and S. Chen, Phys. Rev. B {\bf 104}, 024201 (2021).

\bibitem{teskin} T. Helbig, T. Hofmann, S. Imhof, M. Abdelghany, T. Klessling, L. W. Molenkamp, C. H. Lee, A. Szameit, M. Greiter, and R. Thomale, Nat. Phys. {\bf 16}, 747 (2020).
\bibitem{nhtopoe2}L. Xiao, T. Deng, K. Wang, G. Zhu, Z. Wang, W. Yi, and P. Xue, Nat. Phy. {\bf 16}, 761–766 (2020).
\bibitem{classical1}
	 A. Ghatak, M. Brandenbourger, J. van Wezel, and C. Coulais, Natl. Acad. Sci. USA {\bf 117(47)}, 29561 (2020).
\bibitem{scienceskin} S. Weidemann, M. Kremer, T. Helbig, T. Hofmann, A. Stegmaier, M. Greiter, R. Thomale, and A. Szameit, Science {\bf 368}, 311 (2020).

\bibitem{skinatom} Q. Liang, D. Xie, Z. Dong, H. Li, H. Li, B. Gadway, W. Yi, and B. Yan, Phys. Rev. Lett. {\bf 129}, 070401 (2022).
\bibitem{halfwinding} C. Yin, H. Jiang, L. Li, R. L\"{u}, and S. Chen, Phys. Rev. A {\bf 97}, 052115 (2018).
\bibitem{feshbach} H. Feshbach Ann. Phys. {\bf 5}, 357-390 (1958).
\bibitem{feshbach2} H. Feshbach, Ann. Phys. {\bf 19}, 287-313 (1962).
\bibitem{cohen} C. Cohen-Tannoudji, Cargese Lectures in Physics, Gordon and Breach, New York, (1968).
\bibitem{linresponse1} G. D. Mahan, Many Particle Physics, Third Edition. New York: Plenum. (2000)
\bibitem{linresponse2} Lihong Zhou, Wei Yi, and Xiaoling Cui
Phys. Rev. A {\bf 102}, 043310 (2020).
\end{thebibliography}
\end{document}